\documentstyle[12pt,psfig,epsf]{article}
\begin{document}

\begin{titlepage}
\vspace{-0.5cm}
\title{\begin{flushright}
{\normalsize FTUAM 96/16 }
\end{flushright}
\vspace{1.5cm}
{\Large \bf On compatibility of the Kaluza-Klein 
approach with the COBE experiment \thanks{Extended version of the
contribution to the Proceedings of the Xth Workshop on High Energy 
Physics and Quantum Field Theory,
Zvenigorod (Russia), 20-26 September, 1995.}}}

\author{
Yuri Kubyshin \thanks{On leave of absence from the Institute of Nuclear
Physics, Moscow State University, 119899 Moscow, Russia.}
\thanks{E-mail: kubyshin@delta.ft.uam.es}\\
Departamento de F\'{\i}sica Te\'orica C-XI \\ 
Universidad Aut\'onoma de Madrid \\
Cantoblanco, 28049 Madrid, Spain \\
and \\
 J\'er\^ome Martin  
\thanks{E-mail: jmartin@ccr.jussieu.fr and jmartin@lca1.drp.cbpf.br}\\
Laboratoire de Gravitation et Cosmologies Relativistes \\
Universit\'e Pierre et Marie Curie \\ 
Tour 22/12, Boite courrier 142, 4 place Jussieu \\
75252 Paris Cedex 05, France}

\date{April 15, 1996}

\maketitle

\begin{abstract}

Contributions of primordial gravitational waves to the 
large-angular-scale aniso\-tropies of the cosmic microwave
background radiation
in multidimensional cosmological models (Kaluza-Klein
models) are studied. We derive limits on free parameters of the models
using results of the COBE experiment and other astrophysical data.
It is shown that in
principle there is a room for Kaluza-Klein models as possible candidates
for the description of the Early Universe. However, the
obtained limits are very restrictive. 
Assuming that the anisotropies are mostly due to gravitational waves,
none of the concrete models, analyzed in the article, satisfy them. 
On the other hand, if the contribution of gravitational waves is very
small then a string inspired model is not ruled out.

\end{abstract}
\end{titlepage}

The idea of T. Kaluza and O. Klein \cite{Kaluza-Klein}, 
which was originally proposed as a mean of unification of gravity and 
electromagnetic interactions, has become later an important 
ingredient of many string and supergravity theories (see \cite{Duff} for
reviews) and in this way 
gave rise to Kaluza-Klein cosmological models \cite{KK-cosmology},
\cite{Kolb-book}. Thus, the idea has 
transformed into the physical hypothesis assuming the existence 
of additional (to the three known) spacelike dimensions of the space-time 
with a certain structure. It is quite interesting and it would be 
highly desirable to check if this hypothesis agrees or not with 
available observational data. 
Experimental tests of general relativity in higher dimensions were 
discussed in Ref. \cite{Casas}. 
In this article we use 
the recent observational data on the anisotropies of the cosmic 
microwave background radiation (CMBR) provided by the satellite 
COBE \cite{cobe} to obtain limits on the free parameters describing 
different cosmological scenarios.

We consider Kaluza-Klein cosmological models with the spacetime
$R \times {\cal M}^3_1 \times {\cal M}_2^d$, where the manifold $R
\times {\cal M}_1^3$ represents our four-dimensional Universe, which
is assumed to be a
Friedman-Lema\^\i tre-Robertson-Walker universe with
flat space hypersurfaces, and where the $d$-dimensional manifold
${\cal M}_2^d$ represents
the space of extra dimensions, often called internal space,
which here is assumed to be compact. We
restrict ourselves to the metrics of the form 
\[
g=-{\rm d}t \otimes {\rm d}t +a^2(t) \tilde{g} +b^2(t)\hat{g}, 
\]
where $a(t)$ and $b(t)$ are scale factors of the spaces ${\cal M}_1^3$
and ${\cal M}_2^d$ respectively. 
In the previous equation $\tilde{g}$ is the three-dimensional
metric on ${\cal M}^3_1$ and
$\hat{g}$ is a $d$-dimensional metric on the internal space
${\cal M}^d_2$. 

It is known that 
a multidimensional theory can be always re-written as an effective 
theory in four-dimensional space-time with an infinite 
tower of fields (Kaluza-Klein modes). This includes a zero mode and
massive modes (often called pyrgons) with masses proportional to $1/b$
which form a discrete spectrum 
(since ${\cal M}_2^d$ is compact). From 
the absence of any signals of heavy Kaluza-Klein modes in present day 
high energy experiments it can be 
concluded that $\hbar c/b > (1 \div 10)$ TeV. Studies in Refs. 
\cite{KPW-B} give strong bounds on the time variation of the scale 
$b(t)$ since the time of the nucleosynthesis. They imply that nowadays 
the scale of the internal space is constant 
with very high precision. From these it follows that extra 
dimensions do not produce any significant 
effect now. Contrary to this, in the Early Universe, extra dimensions 
could have played an important role. Indeed, as many cosmological scenarios 
predict, the scales were of comparable size ($a(t) \sim b(t)$) 
and $b(t)$ was changing rather rapidly. Therefore, experiments probing 
the Early Universe can shed light on the issue of the validity of the 
Kaluza-Klein hypothesis. The COBE experiment is the one of this kind:
the temperature anisotropies, measured by COBE, carry 
certain imprints of that early stage of the evolution and, 
in this way, can constrain the topology of the Universe (see
Ref. \cite{topology}) or give an evidence of possible existence of 
extra dimensions. 

The effect we are going to calculate here is in many respects similar
to the temperature anisotropies of the CMBR caused by the inflationary 
expansion of the spatial part of $R\times {\cal M}_1^3$. 
For the case of the four-dimensional
Universe the effect of production of gravitational waves during the
inflationary stage was studied long ago in Ref. \cite{Starobinsky} and 
the effect of distortion of the CMBR by such waves with its consequent
anisotropies was discussed in \cite{Rubakov}. 
A quantum-mechanical mechanism of graviton creation (as amplification
of the zero-point fluctuations) in four dimensions was proposed and 
developed in Ref. \cite{Gr74-90}. 
In Ref. \cite{Gr-PR93} the temperature anisotropies due to the
gravitational wave perturbations generated quantum-mechanically were 
studied and it was shown that a certain 
(although quite narrow) class of four-dimensional inflationary 
scenarios agrees with the observational data \cite{cobe} on the 
anisotropies of the CMBR. (See Refs. \cite{Kolb-book},
\cite{CMBR-review} for reviews on these issues, Refs. \cite{string}
on effects of the gravitational perturbations in string cosmology.) 
In this article we calculate the anisotropies of the temperature of the
CMBR due to the tensor perturbations (gravitational waves) 
within the Kaluza-Klein approach. 
Density perturbations and the spectrum of gravitational waves in 
multidimensional scenarios were studied in \cite{De-GV} - \cite{FCO}. 
An analysis of a class of Kaluza-Klein models and their comparison
with observations of the CMBR were carried out recently in
\cite{FCO}. 

In this article, 
we consider only physical gravitational waves, i.e.
tensor type fluctuations on ${\cal M}^{3}_{1}$, and 
assume that the only
spatial dependence is given by the eigentensors of the Laplacian
on ${\cal M}^3_1$ labelled by the wavenumber $n$ \cite{Ab}, that is 
we retain only the lowest (zero) mode on ${\cal M}^d_2$.
One can show that non-zero (massive) Kaluza-Klein modes do not produce 
any considerable 
contribution to the anisotropies of the CMBR: first, because 
their amplification during inflation is not sufficient and, second, 
because at time of the emission of the CMBR photons they are already heavy 
enough particles. In terms of the conformal 
time $\eta$ the perturbed metric is given by 
\[
ds^{2}= a^{2}(\eta) \left[ - d\eta^{2} + (\delta_{ij} + h_{ij}(\eta,x)) 
dx^{i}dx^{j} \right] + b^{2}(\eta) \hat{g}_{ab}dy^{a} dy^{b}, 
\]
where the indices $i$, $j$ run from $1$ to $3$ and the indices 
$a$, $b$ from $1$ to $d$; the wave can be expressed as
\[
h_{ij}(\eta,x)= \sum_{\{n\}} \frac{\mu_{n}(\eta)}{f(\eta)} 
G_{n;ij}(x), 
\]
where $G_{n;ij}(x)$ are eigentensors of the Laplacian on ${\cal
M}_1^3$ and where the function $f(\eta)$ is defined by 
$f(\eta )\equiv a(\eta) b(\eta)^{d/2}$. The linearized 
Einstein equations with zero energy-momentum tensor give:
\begin{equation}
\label{2-10}
\mu_{n}''(\eta)+ (n^2-\frac{f''(\eta)}{f(\eta)}) \mu_{n}(\eta) =0.
\end{equation}
The physical interpretation of this equation is similar to that 
already given in the four-dimensional case \cite{Gr74-90}, \cite{Gr-CQG93}: 
gravitational waves are parametrically amplified throughout 
the cosmic evolution. In the multidimensional case, 
Eq. (\ref{2-10}) was first considered in Refs. \cite{Ab,De-GV,GG}. 

In a quantum-mechanical treatment \cite{Gr74-90} (see also
\cite{Gr-CQG93}) the
perturbed metric $h_{ij}$ becomes an operator. If we require the amount
of energy to be $\hbar \omega /2$ in each mode, its
general expression can be put under the form:
\[
h_{ij}(\eta ,{\bf x}) = 4\sqrt{\pi
}\frac{l_{Pl}b_{KK}^{d/2}}{f(\eta)}\frac{1}{(2\pi)^{3/2}}\int ^{+\infty
}_{-\infty}{\rm d}^3{\bf n}\sum_{s=1}^{2}p_{ij}^s({\bf n})
\frac{1}{\sqrt{2n}}(c_{\bf n}^s(\eta
)e^{i{\bf n \cdot x}}+c_{\bf n}^{s\dagger}(\eta )e^{-i{\bf n
\cdot x}}).
\]
We used the fact that the multidimensional
gravitational constant $G^{(4+d)}$ is related to the four-dimensional
one $G^{(4)}$ as $G^{(4+d)}=G^{(4)} V_{d}$ with the volume
of the internal space $V_{d}$ evaluated for $b=b_{KK}$, the present day
value of the scale factor of the internal space. The
polarization tensor $p_{ij}^s({\bf n})$ satisfies the relations:
$p_{ij}^sn^j=0$, $p_{ij}^s\delta ^{ij}=0$,
$p_{ij}^sp^{s'ij}=2\delta ^{ss'}$
and $p_{ij}^s(-{\bf n})=p_{ij}^s({\bf n})$. The time evolution of the
operator $h_{ij}(\eta ,{\bf x})$ is determined by the time
evolution of the annihilation and creation operators 
$c_{\bf n}^s$ and $c_{\bf n}^{s\dagger }$ which
obey the Heisenberg equation:
\[
\frac{{\rm d}c_{\bf n}}{{\rm d}\eta } = -i[c_{\bf n},H] ,
\; \; \; \;
\frac{{\rm d}c_{\bf n}^{\dagger}}{{\rm d}\eta } = -i[c_{\bf n}^{\dagger},H] .
\]
The Hamiltonian $H$, providing a description in terms
of travelling waves, is given by:
\begin{equation}
\label{3-21}
H=nc_{\bf n}^{\dagger}c_{\bf n}+nc_{-{\bf n}}^{\dagger}c_{-{\bf n}}+
2\sigma (\eta )c_{\bf n}^{\dagger}c_{-{\bf n}}^{\dagger}+2\sigma ^*(\eta
)c_{\bf n}c_{-{\bf n}},
\end{equation}
where $\sigma(\eta )\equiv if'/(2f)$.
For $d=0$ the expressions of Ref. \cite{Gr-PRL} are recovered.
In the multidimensional case, the second pump field 
$b(\eta)$ appears in the coupling function $\sigma (\eta )$ 
and, as a consequence, the production of gravitons will be 
affected by the dynamics of the internal dimensions. The form of 
the Hamiltonian (\ref{3-21}) explicitly demonstrates
that, while the Universe expands, the initial vacuum state evolves
into a strongly squeezed vacuum 
state with characteristic statistical properties
as discussed in Ref. \cite{Gr-PRL}. The Heisenberg equations are resolved
with the help of the standard Bogoliubov transformations:
$c_{\bf n}(\eta) =u_{n} c_{\bf n}(\eta_{0}) + v_{n} c_{\bf n}^{\dagger}
(\eta_{0})$ and similar one for $c_{\bf n}^{\dagger}(\eta)$. 
Here, $\eta _0$ is the beginning of inflation where the 
normalization is set. Then one can show that the function 
$\mu_{n}(\eta) \equiv u_{n} (\eta) + v_{n}^{*}(\eta)$
obeys the classical equation (\ref{2-10}).

In order to derive bounds on parameters of cosmological models from 
COBE observational data we calculate the angular correlation
function for the temperature variation $(\delta T/T)(\vec{e})$ 
of the CMBR caused by the
cosmological perturbations (Sachs-Wolfe effect \cite{SW}). This
function depends only on the angle $\delta $ between 
the unit vectors $\vec{e}_1$ and $\vec{e}_2$,
pointing out in the directions of observation, and can be expanded
in terms of the Legendre polynomials $P_{l}$ as follows 
\cite{Gr-PRL,Yu-Je}:
\begin{equation}
\label{variance}
<0|\frac{\delta T}{T}(\vec{e}_1)\frac{\delta T}{T}(\vec{e}_2)|0>=
\sum _{l=2}^{\infty }C_lP_l(\cos \delta ). 
\end{equation}
The contribution of the cosmological perturbations to the 
quadrupole moment is given by 
\begin{equation}
Q_{rms-PS}\equiv T_0(5C_2/4\pi)^{1/2}\sim T_{0} h_H/70,
\label{Qh-relation}
\end{equation}
where $T_0=2.7$ K and $h_{H}$ is the characteristic spectral 
component \cite{Gr-PR93} defined by
$ h(n;\eta) = l_{Pl} n |\mu_{n}(\eta)|/a(\eta)$ and evaluated at
$n=n_{H}$, $\eta = \eta_{R}$. 
Here $\eta_{R}$ is the time at which photons of the CMBR were received, 
and $n_{H}=4\pi$ is the wavenumber corresponding to the wavelength 
equal to the present Hubble radius $l_{H}$. 

In this article we consider the following scenario for the
behaviour of the scale factors:

1) Inflationary stage (I-stage):  $\eta_{0}< \eta < \eta_{1} < 0$
\begin{equation}
a(\eta)  =  l_{0}|\eta|^{1+\beta}, \; \; \;
b(\eta)  =  b_{0}|\eta|^{\gamma}. \label{Ia}
\end{equation}

2) Transition stage: $\eta_{1} < \eta < \eta_{2}$
\[
a(\eta) =  l_{0} a_{e} (\eta - \eta_{e}), \; \; \;
b(\eta) =  b_{Tr}(\eta ).
\]

3) Radiation-dominated stage (RD-stage): $\eta_{2} < \eta < \eta_{3} $
\[
a(\eta) =  l_{0} a_{e} (\eta - \eta_{e}), \; \; \;
b(\eta) = b_{KK}.
\]

4) Matter-dominated stage (MD-stage): $\eta > \eta_{3}$
\[
a(\eta) = l_{0} a_{m} (\eta - \eta_{m})^{2}, \; \; \;
b(\eta) = b_{KK}.
\]

The scale factors $a(\eta)$ and $b(\eta)$ and their first derivatives
$a'(\eta)$ and $b'(\eta)$ are assumed to be continuous throughout the
evolution. This gives certain relations between the parameters of the
scenario. 
All models with $1+\beta < 0$ ($\eta$ must be negative in this case) 
describe inflationary expansion of the three-dimensional part of the 
Universe. It can be shown that the case $\beta = -2$ corresponds to
the de Sitter expansion. This scenario is rather 
general since in most of the
known models of Kaluza-Klein cosmology \cite{KRT,GSV}
the behaviour of the scale factors at the inflationary-compactification
stage is of the same type as the one described by Eqs. (\ref{Ia}). 
The transition stage is added in order to assure the continuity 
of $b(\eta)$ and its first derivative. Its behaviour at this stage is
characterized by some reasonable function $b_{Tr}(\eta)$ which describes 
the slowing of the evolution of the scale factor of extra 
dimensions in the process of compactification, that appears in many 
Kaluza-Klein cosmological models (see, for example, \cite{KRT}). 
Thus, a realistic and physically interesting case is when the 
change of $b(\eta)$
during the transition stage is small, namely $r =
(b(\eta_{1})/b_{KK})^{d/2} \sim 1$. The explicit form of $b_{Tr}(\eta)$ 
will not be important for our results. For $\eta > \eta _{2}$ 
the function $b(\eta)$ is taken to be constant. 
This agrees with strong bounds on the time variation
of the scale factor of extra dimensions during the RD- and 
MD-stages obtained in \cite{KPW-B}. 

To calculate the angular variation of the temperature of the CMBR we
need to solve Eq. (\ref{2-10}). The initial conditions on the wave
amplitude, corresponding to the vacuum spectrum of the perturbations
characterized by "a half of the quantum" in each mode, are the following:
$\mu(\eta_{0}) =1$, $\mu'(\eta_{0}) = -in$, where $\eta_{0} < 0$ is such
that $|\eta_{0}| \gg |\eta_{1}|$ \cite{Gr-CQG93}.
Then the solution of Eq. (\ref{2-10}) is equal to

1) I-stage:
\[
\mu(\eta) =(n\eta )^{1/2} A H^{(2)}_{N+\frac{1}{2}}(n\eta ),
\]
where $H^{(2)}_{\nu}(z)$ is the Hankel function of the second kind,
$N\equiv \beta + (\gamma d)/2$ and the constant $A$ given by the expression 
$A=-i\sqrt{\pi/2}\exp [i(n\eta_0-\pi N/2)]$.

2) Transition stage: for waves with long wavelengths 
\[
\mu (\eta ) \sim B_1 \xi(\eta )+B_2 \xi(\eta )\int ^{\eta }
\frac{{\rm d}\eta '}{\xi^2(\eta ')},
\]
where without loss of generality we wrote $b_{Tr}(\eta )$ as 
$b_{Tr}(\eta )=[\xi(\eta )/(a_el_0(\eta -\eta _e))]^{2/d}$, $\xi(\eta
)$ being an arbitrary function (see details in Ref. \cite{Yu-Je}). 

3) RD-stage:
\[
\mu(\eta) =C_{1}e^{-in(\eta -\eta _e)}+C_{2}e^{in(\eta -\eta _e)}.
\]

4) MD-stage:
\[
\mu(\eta) =\sqrt{\frac{\pi
z}{2}}\biggl(D_{1} J_{\frac{3}{2}}(z)+D_{2} J_{-\frac{3}{2}}(z)\biggr),
\]
where $z\equiv n(\eta -\eta _m)$. The coefficients $B_{i}$, $C_{i}$ and
$D_{i}$ $(i=1,2)$ are determined by matching the solution and its first
derivative.

The expressions of $\mu (\eta )$ for the I-, RD- and MD-stages are
exact, whereas the form of $\mu (\eta )$ for the transition stage 
is only valid for long wavelengths, namely for the wave numbers such 
that $n \leq n_{H} = 4\pi$. Such solution is sufficient for 
the calculation of the characteristic spectral component $h_H$. 

To set the scale for $\eta$ it is convenient to choose 
$\eta_{R}- \eta_{m} = 1$.
All realistic cosmological models should account for 
$a(\eta_{E})/a(\eta_{R}) \approx 10^{-3}$ ($\eta_{E}$ 
is the time at which the photons of the CMBR were emitted),
$a(\eta_{3})/a(\eta_{R}) \approx
10^{-4}$ and $a(\eta_{1})/a(\eta_{R}) = k$, where
$3 \cdot 10^{-32} < k < 3 \cdot 10^{-12}$. The lower
bound on $k$ corresponds to the case when the radiation dominated
expansion of the three-dimensional part of the Universe starts at the
Planckian energy densities, whereas the upper one corresponds to
the case when this process starts at the 
nuclear energy densities.
Also we assume that the scale factor $a(\eta)$ has grown
sufficiently during the I-stage: $a(\eta_{1})/a(\eta_{0}) = E$ with
$E \geq e^{70}$ \cite{Guth}. 
We obtain the following expressions for the parameters of the scenario
in terms of $\beta $ and $k$.
\begin{eqnarray}
\eta_{1} &=& 50k(1+\beta ), \nonumber  \\
\eta_{3} &=& 50k\beta +0.5\cdot 10^{-2}, \; \; \;
\eta_{e} = 50k\beta, \; \; \;
\eta_{m} = -0.5\cdot 10^{-2}+50k\beta, \nonumber  \\
a_{e} &=& -(1+\beta )(50k|1+\beta |)^{\beta }, \; \; \;
a_{m} = 50|1+\beta |(50k|1+\beta |)^{\beta }.  \nonumber
\end{eqnarray}
The characteristic scale $l_{0}$ in Eq.
(\ref{Ia}) is given by the relation
\begin{equation}
\frac{l_{Pl}}{l_{0}} = 25 \left( \frac{l_{Pl}}{l_{H}} \right)
     (50k)^{\beta} |1+\beta|^{(1+\beta)},    \label{lPll0}
\end{equation}
where $l_{H} \equiv a^{2}(\eta_{R})/a'(\eta_{R})$ is the present day
Hubble radius. We take it to be equal to $l_{H}=10^{61} l_{Pl}$. 
For the sake of simplicity we also assume that at the beginning of
the inflation the multidimensional Universe was symmetric with
respect to all dimensions, i.e.  $a(\eta_{0})=b(\eta_{0})$. 

Let us now derive restrictions on the parameters of the scenario 
under consideration. One can show that
\begin{equation}
S \equiv \frac{l_{Pl}}{b_{KK}}  =  
   \frac{r^{2/d}}{6} \left( \frac{l_{Pl}}{l_{H}} \right)
   \frac{1}{k/3} E^{\rho}, \label{S-defn}
\end{equation}
where $\rho=(1+\beta-\gamma)/(1+\beta)$. 
The only experimental bound on the size of $b_{KK}$ comes from the fact
that no effects of extra dimensions are observed in high energy
particle experiments. This, apparently, tells us that $\hbar c/b_{KK} >
(1 \div 10) $ TeV. On the other hand the classical description of the
background dynamics can be trusted only if $b_{KK}$ is not much
smaller than $l_{Pl}$. These arguments imply that $S$ in Eq.
(\ref{S-defn}) should belong to the interval $10^{-16}<S<1$. 

To obtain restrictions given by the COBE experimental data we
calculate the characteristic spectral component $h_{H}$ for the
solution $\mu_{n}(\eta)$ for the amplitude of the gravitational 
wave decribed above. We obtain:
\begin{equation}
 h_{H} = 25 \left( \frac{l_{Pl}}{l_{H}} \right)
  |\Psi (M)||1+\beta|^{M} (50k)^{M} n_{H}^{2+M} \Pi ,
  \label{h-final}
\end{equation}
where $M = -1/2 - |1/2 + N|$ and  
$\Psi (M) \equiv -\exp (i\pi M/2) [\Gamma (-1/2-M)]/[\sqrt{2\pi} 
2^{M+1/2}]$. The factor $\Pi $ takes into account the effect of the
transition stage. For $N < -1/2$ it is equal 
to $|1+\beta|/r$, for $N > -1/2$ it is given by 
some complicated expression which in the case of the
transition stage to be short enough reduces to 
$(2\beta+1+d\gamma) r - (1+\beta)/r$. It is easy to show 
that for further analysis the effect of the transition stage 
can be neglected if $r \sim 1$. 
The most recent COBE experimental results give 
$Q_{rms-PS}\sim 15\cdot 10^{-6}$ K \cite{cobe}. 
Since the contribution of tensor perturbations cannot exceed this
value, from Eq. (\ref{Qh-relation}) it follows that 
$h_H \leq 10^{-4}$.

Working out Eqs. (\ref{S-defn}) and (\ref{h-final}), we find that 
\begin{eqnarray}
  \left| N + \frac{1}{2}\right| & = &  - \frac{1}{2} - \frac{57 
                   + \lg h_{H}}{3+\lg (k/3)},  \label{N-bound} \\
  \rho & = & \frac{61 + \lg (k/3) + \lg S}{\lg E},  \label{rho-bound}
\end{eqnarray}
In addition $\beta$ must satisfy $\beta < -1$ to decribe
the inflationary expansion. 

Let us first analyze the case when the observed anisotropies come
mostly from gravitational waves, i.e. $\lg h_{H} = -4$ 
(see also the discussion in Ref. \cite{White}). The 
allowed values of $N$ and $\rho$ are given by the regions bounded by
dashed lines in Fig. 1. 
\begin{figure}[thb]
\vspace{-4cm}
\centerline{\psfig{figure=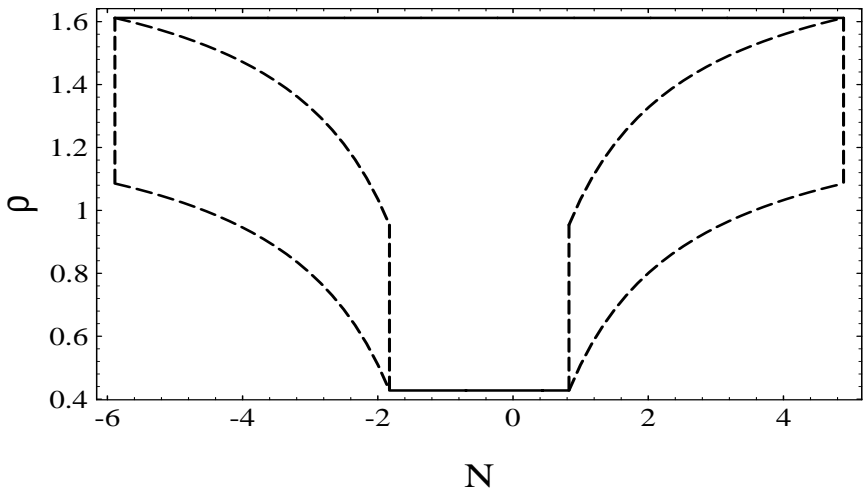,height=15cm,width=14cm}}
\vspace{-4cm}
\caption{The region of values of the parameters
$N$ and $\rho$ given by the equations (9)
and (10) with $E = e^{70}$.  
The subregions bounded by dashed lines correspond 
to the case when the observed anisotropies are given by the gravitational
waves only. $N = - 1/2$ is the symmetry axis of the
whole region.}
\end{figure}
We see that there is a room for
multidimensional cosmological models as candidates for the description
of the Early Universe. However, the limits are rather restrictive, and 
we are unaware of any model of Kaluza-Klein cosmology which agrees with
them. For example,
among the models corresponding to our scenario, one finds that
$\beta=-5/4$, $\gamma = 1/4$ for $d=6$ $[(\rho ,N)=(2,-0.5)]$ 
in the perfect-fluid-dominated
model \cite{ABE}, $\beta=-1.26$, $\gamma=0.22$ in the 
$D=4+d=11$ supergravity with toroidal compactification 
$[(\rho ,N)=(1.85,-0.49)]$ \cite{MN}, $\beta=-14/11$, $\gamma=1/11$
for $d=22$ $[(\rho ,N)=(1.33,-0.27)]$ in the model of 
string-driven inflation \cite{GSV}. It is
easy to check that none of these models satisfy the bounds with $\lg
h_{H} = -4$. 
In these examples the production of the gravitational waves 
is not sufficient to explain the quadrupole moment measured by COBE.

Now let us consider the case when contributions of density and 
rotational perturbations are not neglected, i.e. $\lg h_H \leq
-4$. The allowed values of the parameters are given by the whole region 
presented in Fig. 1. Since $\lg E>30$ the upper horizontal 
line is always below the limit $\rho \sim 49/30=1.63$. As a 
consequence, the perfect fluid model and the supergravity model
do not account for sufficient inflationary growth of 
the three-dimensional part of the Universe within our 
scenario and are still ruled out whereas the string model now 
satisfies the bounds in principle. However, it requires 
$\lg h_H\sim -40$ that means that the contribution of the gravitational 
waves must be extremely small.
  
Taking into account further experimental restrictions (in particular, the 
tensor to scalar quadrupole ratio which is not determined by COBE) will 
allow to make the limits on multidimensional models more restrictive 
thus questioning the very validity
of the Kaluza-Klein hypothesis. 

Details of the calculations as well as results of more complete 
analysis of Kaluza-Klein models will be presented elsewhere
\cite{Yu-Je}. 

\noindent{\large \bf Acknowledgments}

We thank Leonid Grishchuk, Richard Kerner, Valery Rubakov and Gustavo
Yepes for valuable discussions and usefull comments. 
Financial support from M.E.C. (grants SAB94-0087 and SAB95-0224)  
and CIRIT and from the Minist\`ere de la Recherche et 
de l'Enseignement (research grant) and the CBPF (postdoctoral grant) 
are acknowledged.

\end{document}